\documentclass[useAMS,usenatbib,SFB@referee]{mn2e}
\usepackage[dvips]{graphicx}
\input epsf
\title[Star formation triggered by SN explosions]{Star formation triggered by
SN explosions: an application to the stellar association of $\beta$ Pictoris}
\author[C. Melioli et al.]{C. Melioli$^{1}$\thanks{cmelioli@astro.iag.usp.br},
E. M. de Gouveia Dal
Pino$^{1}$\thanks{dalpino@astro.iag.usp.br}, R. de la
Reza$^{2}$\thanks{delareza@on.br}, A. Raga$^{3}$\thanks{raga@nuclecu.unam.mx}\\
$^{1}$ Universidade de S\~ao Paulo, IAG, Rua do Mat\~ao 1226,
Cidade Universit\'aria, S\~ao Paulo 05508-900, Brazil\\
$^{2}$ Observat\'orio Nacional, Rua General Jos\'e Cristino 77, S\~ao
Cristov\~ao, 20921-400 Rio de Janeiro, Brazil\\
$^{3}$ Instituto de Ciencias Nucleares, Universidad Nacional Aut\'onoma de
M\'exico, Ap.P. 70543, 04510 DF, Mexico}

\begin{document}

\date{Accepted ??? ???. Received ??? ???; in original form ??? ??? ???}

\pagerange{\pageref{firstpage}--\pageref{lastpage}} \pubyear{2006}

\maketitle

\label{firstpage}

\begin{abstract}

In the present study, considering the physical
conditions that are relevant in interactions between supernova
remnants (SNRs) and dense molecular clouds for triggering star
formation we have  built a diagram of  SNR radius versus  cloud
density in which the constraints above delineate a shaded zone
where star formation is allowed. We have also performed fully 3-D
radiatively cooling numerical simulations of the impact between SNRs
and clouds under different initial conditions in order to follow
the initial steps of these interactions. We determine the
conditions that may lead either to cloud collapse and star
formation or to complete cloud destruction and find that the
numerical results are consistent with those of the SNR-cloud
density diagram. Finally, we have applied the results above to the
$\beta-$Pictoris stellar association which is composed of low mass
Post-T Tauri stars with an age of 11 Myr. It has been recently
suggested  that its formation could have been triggered by the
shock wave produced by a SN explosion localized at a distance of
about 62 pc that may have occurred either in the Lower Centaurus
Crux (LCC) or in the Upper Centaurus Lupus (UCL) which are both
nearby older subgroups of that association (Ortega and
co-workers). Using the results of the analysis above we have shown
that the suggested origin for the young association at the
proposed distance is plausible only for a very restricted range of
initial conditions for the parent molecular cloud, i.e., a cloud
with a radius of the order of 10 pc and density of the order of
20 cm$^{-3}$ and a temperature of the order of 50$-$100 K.

\end{abstract}


\section{Introduction}

The young Beta Pictoris Association is a nearby unbound moving
group formed by low-mass post-T Tauri stars. Its mean distance to
the Sun is $\sim$ 35 pc (Zuckerman et al. 2001; Torres et al.
2006). Due to its age  ($\sim$ 12  Myr, Zuckerman et al. 2001; or
11.2 $\pm$ 0.3 Myr, Ortega et al. 2002, 2004), this association
has received a lot of attention, since it can be the host of planetary
formation among its star members. Also, this  group contains the
largest number of debris disks and very probably the oldest
classical T Tauri type disk (Torres et al. 2006). Distinctly from
the mechanism usually proposed for star formation in open
clusters, Ortega et al. (2002, 2004) have suggested that a type II
supernova could have triggered the formation of this loose group
of stars. That could also be the case for the younger association
of TW Hya 8 Myr ago (de la Reza et al. 2006).

A star formation process starts when a pressure-bounded,
self-gravitating cloud or clump becomes gravitationally unstable.
In the classical theory developed by Jeans, an instability occurs
when the gravitational attraction overcomes the combined action of
all dispersive and resistive forces. The simplest case is a
system in virial equilibrium where only  the presence of the
potential energy is considered. If the potential energy is greater
than twice the total kinetic energy, the system collapses, while
in the opposite case, it expands. This consideration can be
extended by including any relevant additional physical forces acting on
the gas.

It has long been known that essentially all present day star formation
takes place in giant molecular clouds (GMCs, e.g. Blitz 1993, Williams,
Blitz, \& McKee 2000), so that it is vital to understand the
properties, dynamical evolution and fragmentation of these clouds
in order to understand star formation.

There are several possible ways of driving a cloud or clump to collapse,
but at the present time is not well understood how a
self-gravitating cloud becomes gravitationally unstable. GMCs are
observed to contain a wealth of structures on all length scales
with highly supersonic motions (Larson 1981; Blitz \& Williams
1999; Elmegreen \& Scalo 2004). Based on numerical simulations, a
number of authors have argued that it is these supersonic motions,
maintained by internal or external driving mechanisms, that induce
the observed density inhomogeneities in the gas (Mac Low \&
Klessen 2004; Elmegreen \& Scalo 2004), and that it is therefore
the supersonic motions that drive star formation. Suggested
candidates for an internal driving mechanism include feedback from
low-mass star formation  although GMCs with and without star
formation have similar kinematic properties (Williams, Blitz \&
McKee 2000). External candidates include galactic spiral shocks
(Roberts 1969, Bonnell et al. 2006) and supernova and superbubble
shocks (Wada \& Norman 2001; Elemegreen \& Scalo 2004). All of these
processes seem to have sufficient energy to explain the kinematics
of the ISM and can generate the observed velocity
dispersion-sizescale relation (Kornreich \& Scalo 2000). Other
mechanisms, such as protostellar winds and jets, magnetorotational
instabilities, expansion of H II regions and fluctuations in the
UV field apparently inject energy into the ambient medium at a
rate which is about an order of magnitude lower than the energy
that is required to explain the random motions of the ISM at
several scales.

Here we focus on one of these driving mechanisms and investigate
the possibility that a SN that exploded in the recent past  produced a
shock front that compressed a molecular cloud and induced the
formation of the $\beta$ Pictoris association,
as suggested by Ortega et al. (2002, 2004). \footnote{We notice that a 
more general study of the role that SN
explosions play on the process of structure and star formation
and on the generation of turbulence in the ISM of normal and
starburst galaxies will be presented elsewhere (Melioli et al. 2006).}

In \S 2, we consider the equations that are relevant for a SNR
expansion and the equations that describe the interaction between
an expanding SNR and a molecular cloud. In \S 3, we obtain
analytically a set of constraints from these interactions that may
lead to star formation and build a diagram where these constraints
approximately delineate a zone in the parameter space which is
appropriate for star formation. In \S 4, we apply these analytical
results to the physical conditions that could have led to the
formation of the $\beta$ Pictoris association induced by a
SNR-cloud interaction and  also describe radiatively cooling 3-D
hydrodynamical simulations of the interactions between a SNR and a
molecular cloud considering initial conditions which are
appropriate to the $\beta$ Pictoris environment and find that the
results confirm those obtained from the analytical study. In \S 5
we draw our conclusions.

\section{SNR evolution and shock wave-cloud interactions}

\subsection{Conditions of the SNR}

A type II supernova explosion generates a spherical shock wave
that sweeps the interstellar medium (ISM), leading to the
formation of a supernova remnant (SNR). The  interaction between
a SNR and a cloud may compress the gas sufficiently to drive the
collapse of the cloud. The kinetic energy associated to a SN event
is of the order of $10^{51}$ erg, the ejected mass into the ISM is
$M_{ej} \simeq 10$ M$_{\odot}$ and its  terminal velocity is $\sim
10^4$ km s$^{-1}$. This ejected mass will expand at nearly
constant velocity until it encounters a comparable mass of
ambient medium. This occurs at a time $t_{sh}$ which determines
the onset of the SNR formation (e.g., McCray 1985):

\begin{equation}
t_{sh} =  {200 \over {n^{1/3}}} \left({M_{ej} \over M_{\odot}} \right)^{1/3}
\ \ \ \ {\rm yr}
\end{equation}
\noindent
where $n$ is the ambient number density in cm$^{-3}$. The SNR evolution is
characterized by two phases: an adiabatic (or {\it Sedov-Taylor})
phase and a radiative phase. In the adiabatic phase the radius, $R_{snr}$,
and the expansion velocity , $v_{snr}$, of the SNR are, respectively:

\begin{equation}
R_{snr}(t) \sim 13 \ \left({{E_{51}} \over n} \right)^{1/5} \ t_4^{2/5} \ \ \
{\rm pc}
\end{equation}

\begin{equation}
v_{snr}(t) \sim 508 \ \left({{E_{51}} \over n} \right)^{1/5} \ t_4^{-3/5} \ \
\ {\rm km/s}
\end{equation}
\noindent where $E_{51}$ is the initial SN energy in units of
$10^{51}$ erg and $t_4$ is the time in units of $10^4$ yr. Using
Eqs. 1 and 2, we obtain:

\begin{equation}
v_{snr}(R) \sim 68 \ \left({{E_{51}} \over n} \right)^{0.5} \ {1
\over {R_{snr,50}^{1.5}}} \ \ \ {\rm km/s}
\end{equation}
\noindent
where $R_{snr,50}$ is the radius of the SNR in units of 50 pc.
The effects of the radiative losses become important at a time (McCray 1985):

\begin{equation}
t_{cool} = 3 \times 10^4 \ {E_{51}}^{0.22} \ n^{-0.55} \ \ \ \
{\rm yr}
\end{equation}
\noindent 
after which the SNR enters the radiative phase and its
evolution in the ISM is then described by:

\begin{equation}
R_{snr}(t) \sim 19 \ {{E_{51}^{0.23}} \over {n^{0.26}}} \
t_4^{2/7} \ \ \ {\rm pc}
\end{equation}

\begin{equation}
v_{snr}(t) \sim 530 \ {{E_{51}^{0.23}} \over {n^{0.26}}} \
t_4^{-5/7} \ \ \ {\rm km/s}
\end{equation}
\noindent
From  Eqs. 6 and 7 we obtain:

\begin{equation}
v_{snr}(R) \sim 47 \ {{E_{51}^{0.8}} \over
{n^{0.91} \ R_{snr,50}^{5/2}}} \ \ \ {\rm km/s}
\end{equation}
\noindent
When the internal pressure of the SNR becomes comparable
to the ISM pressure, it stalls, fragments and the hot gas
that fills the SNR begins to mix with the ISM. Assuming that the
radiative losses of the hot gas are negligible, this phase is
expected to occur when:

\begin{equation}
R_{snr} \sim 56 \ T_4^{-0.2} \ {E_{51}}^{0.12} \ n^{-0.37} \ \ \ {\rm pc}
\end{equation}
\noindent 
where $T_4$ is the ISM temperature in units of $10^4$ K.

\subsection{Conditions for a SNR-cloud interaction}
Molecular clouds are  dominated by molecular H$_2$ because they
are opaque to the UV radiation that elsewhere dissociates the
molecules. The heating and cooling processes in molecular clouds
are due mainly to the presence of molecules composed of heavier elements
such as carbon,
nitrogen, and oxygen. Emission line observations reveal clumps
and filaments on all scales. While the GMCs have masses of $10^5$
to $10^6$ M$_{\odot}$ and extend over a few $\sim$ 10 pc, the
smallest embedded structures are protostellar cores with masses of
a few solar masses or less and sizes smaller than 0.1 pc.
The cloud temperature typically ranges from 10 K to 100 K and only
in the presence of a photo-ionizing source it increases to $10^4$ K.
If we assume a homogeneous cloud with constant density and temperature,
we can describe its evolution after an
interaction with a SNR shell moving at a supersonic velocity.

After the impact, an internal forward shock propagates
into the cloud  with a velocity $v_{cs}$. The ram pressure of the
blast wave, $\sim n_{sh} v_{snr}^2$, must be comparable to the
ram pressure behind the shock in the cloud, $\sim n_{c} v_{cs}^2$
and this results in:

\begin{equation}
v_{cs,A} \sim v_{snr} \ \left({{n_{sh}} \over {n_c}}\right)^{0.5} = {{43 \
E_{51}^{0.5}} \over {R_{snr,50}^{1.5} \ n_{c,10}^{0.5}}} \ \ \ {\rm km/s}
\end{equation}
\noindent 
in the adiabatic case, where $n_{c,10}$ is the cloud
density in units of 10 cm$^{-3}$ and where we have assumed $n_{sh}
= 4n$ (as appropriate for a strong adiabatic shock), and

\begin{equation}
v_{cs,R} \sim f_{10}^{0.5} \ {{46 \ E_{51}^{0.8}} \over
{R_{snr,50}^{5/2} \ n_{c,10}^{0.5} \ n^{0.41}}} \ \ \ {\rm km/s}
\end{equation}
\noindent
in the radiative case, where $f_{10}$ is the density contrast between the shell
and the ISM density, in units of 10.

The equations above are valid for a planar shock. For spherical
cloud-SNR interactions, the effects of curvature in the shock
should be considered. The instantaneous velocity of the shocked
gas moving towards the center of the cloud is only a fraction of
the SNR velocity and depends on the density contrast $\chi$
between the shell and the cloud (as in the planar shock case) and
the angle $\gamma$ between the SNR velocity vector and the line
that links the center of the cloud and the instantaneous contact
point between the cloud and the SNR (see Figure 1). When the SNR
touches the cloud, these two lines are coincident ($\gamma$ = 0)
and $v_{cs}={\chi}^{0.5}V_{snr}$. Later, when the SNR crosses the
center of the cloud, $\gamma = \pi /2$, and $v_{cs}=0$. The
average value of the velocity integrated over the SNR crossing
time, $t_{c,snr}$ is:

\begin{equation}
{\hat{v}}_{cs} = v_{snr} \ \left({{n_{sh}} \over {n_c}}\right)^{0.5} \ {1
\over t_{c,snr}} \ \int^{t_{c,snr}}_0{cos \gamma(t) \ dt}
\end{equation}
\noindent
where $t_{c,snr}=2R_c/v_{snr}$.

\begin{figure}
\centering
\epsfxsize=8cm
\epsfbox{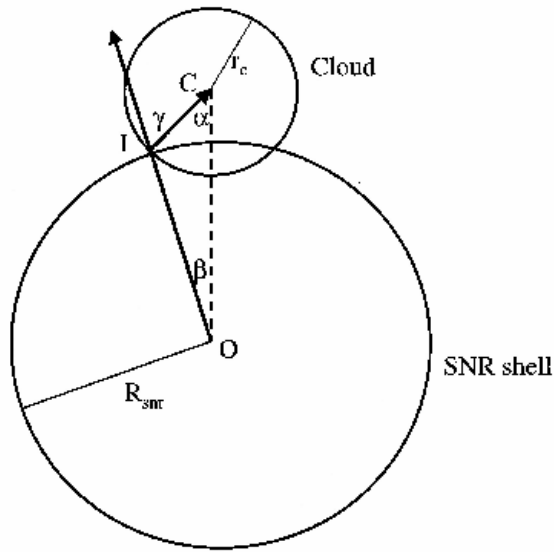}
\caption{Schematic diagram of the
interaction between a SNR and a cloud. The SNR expands and impacts
the cloud. The angles $\alpha$, $\beta$ and $\gamma$ are functions
of the time, the SNR velocity and the cloud and SNR radii, as
indicated by the equations of the text.}
\end{figure}
\noindent From Figure 1 we find that:
$$\gamma(t) = \alpha(t) + \beta(t)$$
where:
$$\alpha(t) = cos^{-1}(1-{{v_{snr}t} \over r_c})$$
$$\beta(t) = sen^{-1}[{{(2v_{snr}r_{c}t-v_{snr}^2t^2)^{0.5}} \over
{R_{snr}}} ]$$
\noindent 
and $r_c$ is the cloud radius expressed
in pc.

Figure 2 depicts the plot of the integral term ($I=1/t_{c,snr}
\int^{t_{c,snr}}_0{cos \gamma(t) \ dt}$) of Eq. (15) as a function
of the ratio $R_{snr}/r_c$.

\begin{figure}
\centering
\epsfxsize=8cm
\epsfbox{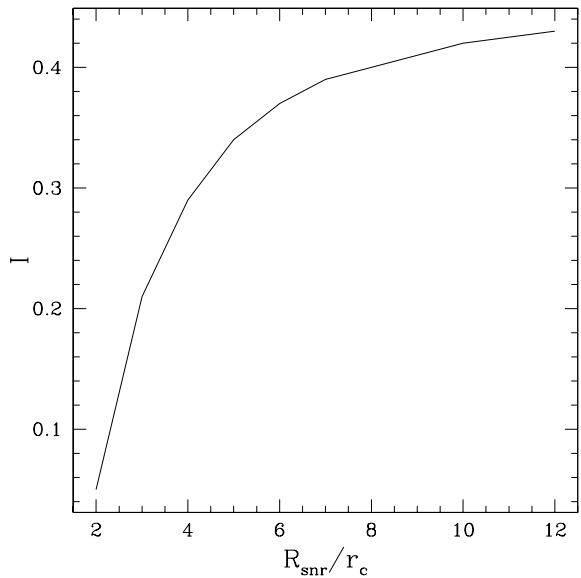}
\caption{Values of I for different ratios $R_{SNR}/r_c$}
\end{figure}
\noindent
For example, for a SNR with $R_{snr}=$ 50 pc and a cloud with
$r_c=$ 10 pc, we find:

\begin{equation}
{\hat{v}}_{cs} = 0.34 \ ({{n_{bw}} \over {n_c}})^{0.5} \ v_{snr}
\end{equation}
\noindent
and the internal shock wave will cross the cloud in a time $t_{cc}$:

\begin{equation}
t_{cc,A} \sim 7 \times 10^5 \ {{n_{c,10}^{0.5} \ r_{c,10} \ R_{snr,50}^{1.5}}
\over {I_5 \ E_{51}^{0.5}}} \ \ \ \ {\rm yr}
\end{equation}
\noindent 
in the case of a SNR in the adiabatic phase, where
$r_{c,10}$ is the cloud
radius in units of 10 pc and $I_5$ is the integral $I$ computed
for a ratio $R_{snr}/r_c=5$. In the case of a SNR in the
radiative phase, we obtain:

\begin{equation}
t_{cc,R} \sim 7 \times 10^5 {{r_{c,10} \ R_{snr,50}^{2.5} \ n_{c,10}^{0.5} \
n^{0.41}} \over {f_{10}^{0.5} \ E_{51}^{0.8}}}  \ \ \ \ {\rm yr.}
\end{equation}
\noindent 
After this time, the shocked gas reaches the center of
the cloud, rebounds and, in the absence of the effects of a
gravitational field, the compressed cloud may start a re-expansion
phase.
The equations above imply that when the SNR interacts
with a cloud the Mach number of the forward shock propagating into
the cloud, is:

\begin{equation}
M_A = 14 {{E_{51}^{0.5}} \over {T_{c,100}^{0.5} \
R_{snr,50}^{1.5} \ n_{c,10}^{0.5}}} \ I_5
\end{equation}
\noindent
if the SNR at the epoch of the interaction is yet in the
adiabatic phase, and

\begin{equation}
M_R = 44 \ {{f_{10} \ E_{51}^{0.8} \ I_5} \over {n_{c,10}^{0.5} \
T_{c,100}^{0.5} \ R_{snr,50}^{2.5} \ n^{0.41}}}
\end{equation}
\noindent 
for an interacting SNR in the radiative phase, where $T_{c,100}$
is the cloud temperature in units of 100 K.
\footnote{We note that the choice of the set of equations above
that are appropriate to describe a given cloud-SNR interaction,
with the determination of whether the SNR is in the adiabatic or
in the radiative phase, will strictly depend upon the initial SN
energy and the ambient medium conditions at the time when the
interaction initiates, which are established by eqs. (2) to (8).}

Depending on the physical conditions of the cloud, the propagation
of the shock front into it can be either adiabatic or
radiative. For the shocked gas at temperatures T $\le 10^4$ K, we
find that the radiative cooling time is shorter than the crushing
time (see, e.g., Melioli, de Gouveia Dal Pino \& Raga 2005) and therefore,
we can assume that the forward shock wave propagating into the cloud is
approximately radiative.
The Rankine-Hugoniot
relations for a radiative strong shock (with M $\ge$ 10), are:

\begin{equation}
T_{c,sh}=T_c
\end{equation}

\begin{equation}
n_{c,sh}=M^2 \ n_c
\end{equation}
\noindent 
where $n_{c,sh}$ and $T_{c,sh}$ are the density and the
temperature of the cloud shocked gas, respectively. Using Eqs. (16)
and (17), we can then estimate the value for the cloud gas density
after the interaction with a SNR, that is:

\begin{equation}
n_{c,sh,A} \sim {1800 \over {{R_{snr,50}^3}}} \ {{E_{51} \ I_5^2} \over
{T_{c,100}}} \ \ \ \ \ {\rm cm^{-3}}
\end{equation}
\noindent 
for a cloud shocked by a SNR in the adiabatic phase, and

\begin{equation}
n_{c,sh,R} \sim {2300 \over {{R_{snr,50}^5}}} \ {{E_{51}^{1.6} \ I_5^2 \
f_{10}} \over {T_{c,100} \ n^{0.82}}} \ \ \ \ \ {\rm cm^{-3}}
\end{equation}
\noindent 
for a cloud shocked by a SNR in the radiative phase.

\section{The conditions for star Formation}
\subsection{The Jeans mass constraint}
In principle, in order to find out the conditions for star
formation, the cloud (or clump) mass should be larger than the
Jeans mass (Jeans, 1902) which, in the absence of magnetic fields,
may be written as:

\begin{equation}
m_J \simeq 1.4 \times 10^{-10} \ {{T^{1.5}} \over {\rho^{0.5}}} \
\ \ \ {\rm M_{\odot}}
\end{equation}
\noindent 
where $\rho$ is the cloud density  in g cm$^{-3}$ and $T$ is the
temperature in K. For a
cloud shocked by a SNR, this condition may be approximately
expressed as (using Eqs. 18, to 21):

\begin{equation}
m_{J,A} \simeq 2200 \ {{T_{c,100}^2 \ R_{snr,50}^{1.5}} \over {I_5
\ E_{51}^{0.5}}} \ \ \ \ {\rm M_{\odot}}
\end{equation}
\noindent 
if the interacting SNR is still in the adiabatic phase,
or

\begin{equation}
m_{J,R} \simeq 2000 \ {{T_{c,100}^2 \ R_{snr,50}^{2.5} \ n^{0.41}}
\over {I_5 \ f_{10}^{0.5} \ E_{51}^{0.8}}} \ \ \ \ {\rm M_{\odot}}
\end{equation}
\noindent 
if the SNR is already in the radiative phase. These
conditions may  be also expressed in terms of the cloud radius:

\begin{equation}
r_{c,A} \ge 12 {{T_{c,100}^{2/3} \ R_{snr,50}^{0.5}} \over {I_5^{1/3} \
n_{c,10}^{1/3} \ E_{51}^{0.17}}} \ \ \ {\rm pc}
\end{equation}
\noindent 
for a cloud interaction with an adiabatic SNR, or

\begin{equation}
r_{c,R} \ge 11.4 {{T_{c,100}^{2/3} \ R_{snr,50}^{0.83} \ n^{0.14}}
\over {I_5^{1/3} \ f_{10}^{0.17} \ E_{51}^{0.27}}} \ \ \ {\rm pc}
\end{equation}
\noindent
for an interaction with a radiative SNR.
Under the conditions above the shocked cloud is gravitationally unstable
and may, in principle, start to collapse.

\subsection{Cloud destruction constraint due to strong SNR impact}

On the other hand, another effect may occur. If the SNR-cloud
interaction is too strong the cloud can be completely destroyed.
To check the conditions for this situation, we should compare the
cloud gravitational free-fall timescale, $t_{ff}$, with the
destruction timescale due to a strong impact, $t_d$. In order to
have collapse, the gravitationally unstable mode (with typical
time $t_{un}$) must grow fast enough to become nonlinear within
the time scale of the cloud-SNR shock interaction (Nakamura et al.
2005). Previous numerical hydrodynamical simulations performed by
several authors under an adiabatic approximation (see, e.g.,
Klein, McKee \& Colella 1994; Poludnenko, Frank \& Blackman 2002)
have shown that the density of the cloud may drop by a factor 2 in
1.5-2 $t_{cc}$ after the impact with a strong shock front. In the
presence of radiative cooling of the shocked gas, this density
drops by the same factor in $\sim$ 4-6 $t_{cc}$ (Melioli, de
Gouveia Dal Pino \& Raga 2005). Thus, in order to have collapse,
we assume that should be $t_{un} \le 3t_{cc}$. 
This implies a Mach number:

\begin{equation}
M \le 14 \ \large{(}{{n_{c,10}} \over {T_{c,100}}}\large{)}^{1.16} \ r_{c,10}^3
\end{equation}
\noindent 
or

\begin{equation}
R_{snr,A} \ge 34 \ {{E_{51}^{0.33} \ T_{c,100}^{0.44} \ I_5} \over {n_{c,10} \
r_{c,10}^2}} \ \ \ \ {\rm pc}
\end{equation}
\noindent 
for the interaction with a SNR in the adiabatic phase,
and

\begin{equation}
R_{snr,R} \ge 50 \ {{E_{51}^{0.33} \ f_{10}^{0.2} \ T_{c,100}^{0.26} \
I_5^{0.4}} \over {n_{c,10}^{0.7} \ n^{0.17} \ r_{c,10}^{1.2}}} \ \ \ \ {\rm pc}
\end{equation}
\noindent 
for the interaction with a SNR in the radiative phase.

Eqs. (22) to (26) and Eqs. (28) and (29) indicate  that a SNR
interacting with a cloud should on one side have sufficient
energy to induce cloud collapse and on the other side be evolved
enough in order to not destroy it. These conditions depend on
the radius, density and less strongly on the temperature of
the cloud, and also on the ISM density and the initial SN energy.

\subsection{Penetration extent of the SNR shock front into the cloud}

Besides the constraints above, another effect upon the shock
should be considered. When the shock wave starts propagating into
the cloud, it is decelerated not only by the cloud material, but
also by the decay of its pressure due to radiative cooling. This
is a non linear situation because the radiative losses depend on
the shocked gas temperature, that in turn depends on the shock
velocity and on the shocked cloud density, that in turn is
sensitive to the radiative losses. However, using energy
conservation arguments, we can at least estimate  the approximate
time  at which  the shock  propagation will stop within the cloud.
At this time ($t_{st}$),  the velocity of the shocked gas in
the cloud goes to zero and its pressure must balance that of the
upstream un-shocked cloud material.
From energy conservation, we find
approximately that:

\begin{eqnarray}
{3 \over 2}  n_{c,sh}(0) k_b T_{c,sh}(0) + {1 \over 2} \mu m_H
n_{c,sh}(0)  v_{c,sh}(0)^2 \simeq
\nonumber
\end{eqnarray}
\begin{equation}
{3 \over 2} n_{c,sh}(t_{st})
k_b T_{c,sh}(t_{st})+\Lambda [T_{c,sh}(t_{st})] \
n_{c,sh}(t_{st})^2 \ t_{st}
\end{equation}
\noindent 
where the LHS is the total energy behind the
shock into the cloud, immediately after the impact and the RHS
is the total energy behind the shock after a time
$t_{st}$ when it stops. $\Lambda(T)$ is the shocked gas
radiative cooling function (which can be approximated by that of
an optically thin gas; see e.g., Dalgarno \& McCray 1972).
The initial temperature and density of the shocked cloud material are
approximately given by the adiabatic values. At $t_{st}$,
$n_{c,sh}(t_{st})$ and $T_{c,sh}(t_{st})$ are obtained from
the pressure balance above between shocked and unshocked material
in the cloud (with the final density approximately equal to the
cloud density). The substitution of these relations into the
equation above results:

\begin{equation}
t_{st} \simeq {9 \over 16} {{\mu m_H} \over {n_c \Lambda}}
v_{sh,c}^2\ \ \ {\rm s.}
\end{equation}
\noindent

This time is used to compute the maximum distance that the shock
front (initiated by a given SNR) can travel into the cloud before
being stopped. This distance is then compared with the radius of
the cloud in order to establish the maximum size (that is, the
minimum energy) that the SNR should have in order to generate a
shock wave able to sweep all of the cloud before being stalled:

\begin{equation}
R_{snr} \le 75 {{E_{51}^{0.33} I_5^{0.66}} \over
{(r_{c,10} \Lambda_{27})^{2/9} n_{c,10}^{0.5}}} \ \ \ {\rm pc}
\end{equation}
\noindent
where $\Lambda_{27}$ is the cooling function in units of $10^{-27}$.

This constraint together with the other limits inferred from Eqs.
(9), (25), and (28) have been plotted together in Figure 3 that
shows the supernova radius as a function of the initial
(unshocked) cloud density for different values of the cloud
radius. We note that these different constraints  provide a
restricted shaded zone where the combination of these parameters
creates appropriate conditions for  star formation. The Jeans mass
constraint derived from Eq. (25) for an interaction with an
adiabatic SNR (dashed line) and the cloud shock deceleration
constraint derived from Eqs. (30) to (32) (dotted line) determine
upper limits for the SNR radius, while the condition for complete
cloud destruction after an encounter with an adiabatic SNR derived
from  Eq. (28) (solid line) determines a lower limit for the SNR
radius. We have taken a SNR in the adiabatic
phase because it has more stored energy than one with the same
characteristics in the radiative phase.
Only cloud-SNR
interactions with initial physical conditions ($r_c$, $n_c$ and
$R_{snr}$) lying within the shaded region of the figure (between
the solid, dotted and dashed lines) may lead to a process of star
formation. \footnote{We notice that the diagrams of Figure 3
have been built for a fixed temperature $T_c = 100$ K. According to Eqs.
(25), (28), and (30-32), they are not very sensitive to this
parameter, except for the Jeans mass constraint given by Eq. (25), which implies that the upper limit for $R_{snr} \propto T^{-1.3}$. We further 
notice that a cloud with a temperature in the range of 10 to 50 K and a 
radius larger than 10 pc is already Jeans unstable over a large range of densities (≥ 5 cm$^-3$) and  does not require (according to the equations above) an interaction with a shock wave to  trigger star formation.}

\begin{figure}
\centering
\epsfxsize=5cm
\epsfbox{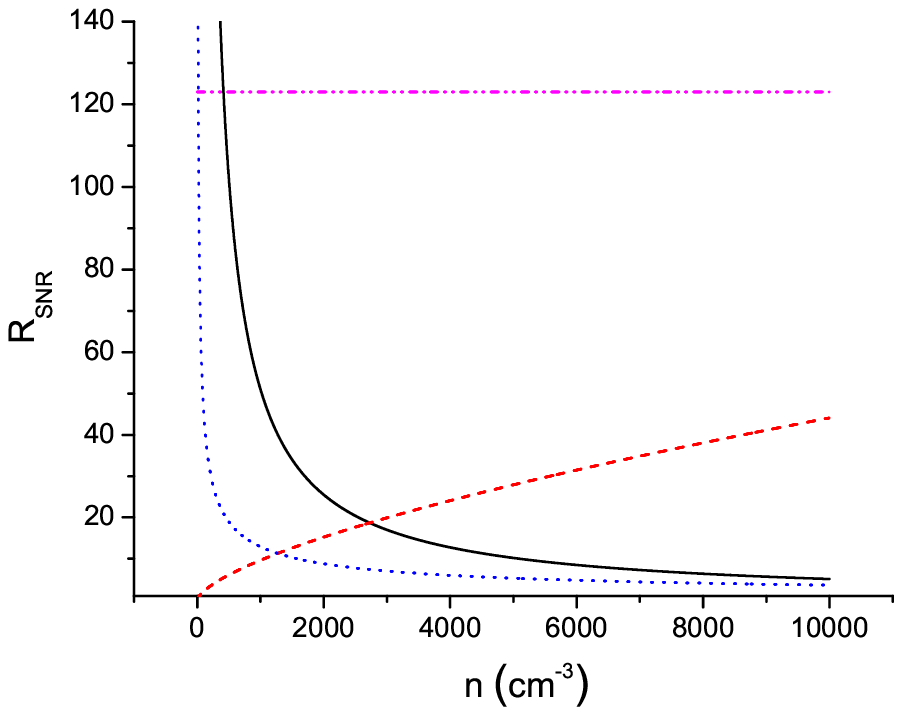}
\epsfxsize=5cm
\epsfbox{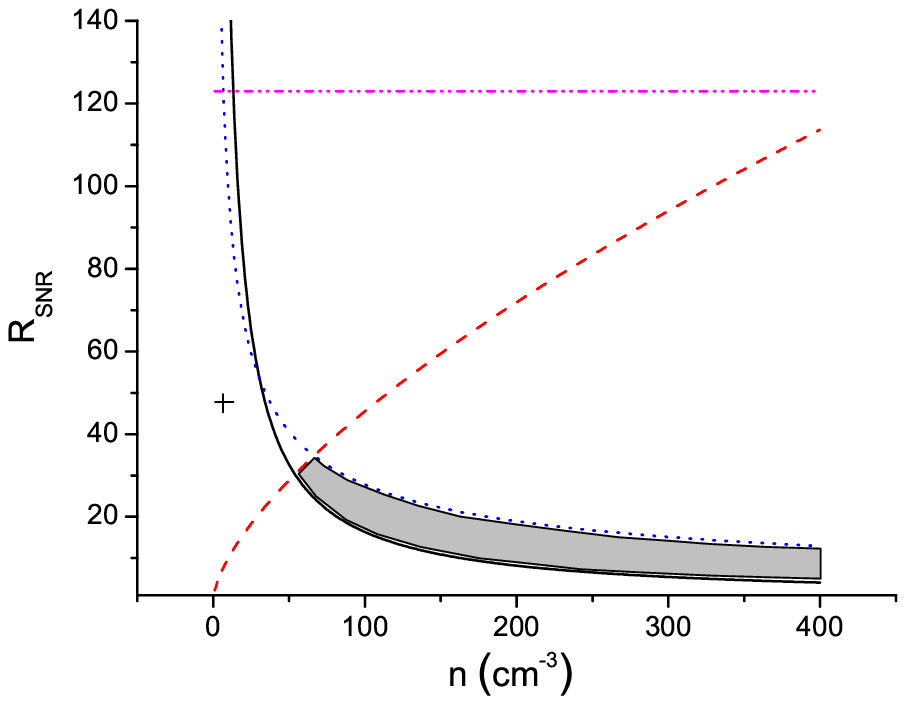}
\epsfxsize=5cm
\epsfbox{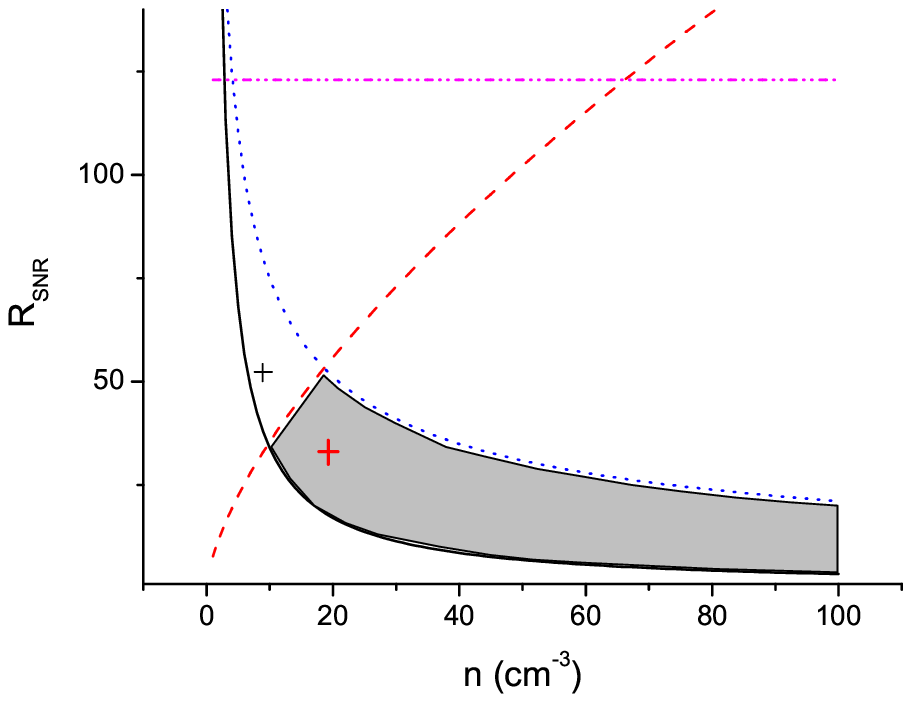}
\epsfxsize=5cm
\epsfbox{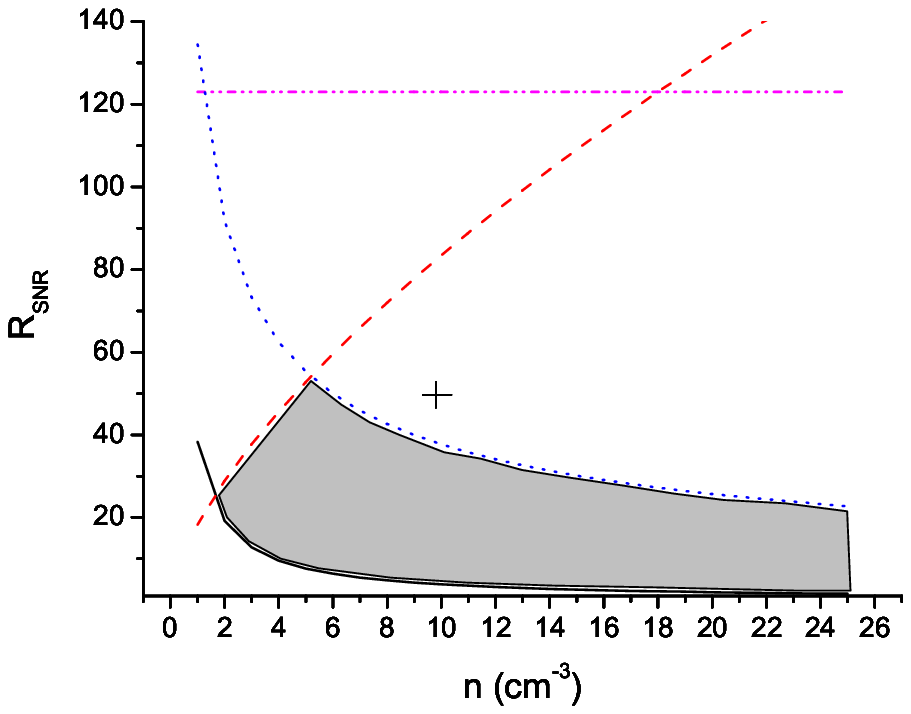}
\caption{Constraints on the SNR radius  versus cloud density for 4
different cloud radius. Top-left panel: $r_c$ = 1 pc; top-right
panel: $r_c$ = 5 pc; bottom-left panel: $r_c$ = 10 pc; and
bottom-right panel: $r_c$ = 20 pc. Solid (black) line: upper limit
for complete cloud destruction after an encounter with an
adiabatic SNR derived from Eq. 28; dashed (red) line: upper limit
for the cloud to reach the Jeans mass derived from Eq. 25 for an
interaction with an adiabatic SNR; dotted (blue) line: upper limit
for the shock front to travel into the entire cloud before being
decelerated to subsonic velocities derived from Eqs. 30 to 34;
dotted-dashed (pink) line: maximum radius reached by a SNR in an
ISM with density n=0.05 cm$^{-3}$ and temperature T=$10^4$ K
derived from Eq. 9. The shaded area defines the region where star
formation can be induced by a SNR-cloud  interaction (between the
solid, dashed and dotted lines). The crosses in the panels
indicate the initial conditions assumed for the clouds in the
numerical simulations described in Section 4.2}
\end{figure}

\section{An application to $\beta$ Pictoris association}

\subsection{SNR-cloud interaction: analytical approach}

In order to apply the simple analytical study described above to
the young moving group of low-mass Post-T Tauri stars  of $\beta$
Pictoris (or BPMG) we assume, as in Ortega et al. (2004), that a
type II SN exploded in the past either in the Lower Centaurus Crux
(LCC) or in the Upper Centaurus Lupus (UCL), both subgroups of the
OB Sco-Cen association, and triggered the formation of BPMG. The
potential position of the SN was obtained by tracing back the past
position of the runaway star HIP 46950. Those authors estimated
that the nearest possible past position of the SN with respect to
the calculated stellar dynamical birthplace of BPMG was of 87 pc.
Nevertheless, this distance is somewhat affected by the
uncertainty of the radial velocity of the runaway star used (35.0
to 10 km/s; Hoogerwerf et al. 2001) and here we will adopt a more
conservative value of $\sim 60$ pc for the SN-cloud distance.
Also, taking an ISM number density of 0.05 cm$^{-3}$ and initial
cloud parameters $r_c \simeq$ 10 pc, $n_c \simeq$ 10 cm$^{-3}$,
and $T_c \simeq$ 100 K, which are typical for GMCs (e.g., Cernicharo 
1991; Mac Low \& Klessen 2004), we find that, for a SN energy of 
$10^{51}$ erg, the SNR-cloud interaction would have occurred after 
$\sim 7 \times 10^4$ yr of the SN explosion (Eq. 2). 
With a SNR velocity $v_{snr} \simeq$ 280 km/s (Eq. (3)), the shock 
velocity into the cloud would be $\hat v_{cs}$, $\sim$ 13 km/s (Eq. 12) 
and after a time $t_{cc} \sim 7 \times 10^5$ yr (Eq. 14), the cloud 
would be compressed developing a core with shocked material with a 
density of $n_{c,sh,A} \simeq 1600$ cm$^{-3}$ (Eq. (20)) and temperature
$T_{c,sh} \simeq$ 100 K (Eq. (18)). However, in order to reach the
Jeans mass, the initial radius of the cloud should be (according
to Eq. 25), $r_c \ge 12.5 \ \ \ {\rm pc}$, which is larger than the
initial radius above and therefore, it indicates that the proposed
SNR-cloud interaction is unable to develop a gravitationally
unstable system, at least under the initial conditions
above. Moreover, we notice that the shock front into the cloud
should be highly decelerated under such conditions, and that its
average velocity would be $\sim v_{c,sh}/3 \sim 4$ km s$^{-1}$.
Also, we estimate that after the interaction with the SNR, the
final density of the shocked cloud material will drop to $\sim
n_{c,sh}/9 \sim 170$ cm$^{-3}$, and in order to reach the Jeans
mass condition the initial radius of the cloud should be $r_c \sim
18$ pc.
On the other hand, back-tracing of the motion of the
Beta-Pictoris members indicate that they may have dispersed from a
region of radius $\sim 15-24$ pc (Song et al. 2003, Ortega et al.
2002, 2004). If we then take for the progenitor cloud an initial
radius of $\sim $18 pc or larger as indicated above, we find that
the forward shock wave is unable to sweep the entire cloud before
stalling (Eq. 32), and therefore, also in these cases, the
SNR-cloud interaction would  be not efficient enough to develop a
process of star formation.

\subsection{SNR-cloud interaction: numerical simulations}

In order to check the semi-analytical estimates above, we have
also performed fully 3-D hydrodynamical radiatively cooling
simulations employing an unmagnetized modified version of the
adaptative grid code YGUAZU (Raga, Navarro-Gonz\'alez \&
Villagr\'an-Muniz 2000; Raga et al. 2002; Masciadri et al. 2002,
Melioli, de Gouveia Dal Pino \& Raga 2005) which solve
the gas dynamical equations together with a set of continuity
equations for several atomic/ionic species (see details in Raga et
al. 2002).
The computational box has dimensions 102 pc $\times$ 56 pc $\times$ 56 pc,
corresponding to 256 $\times$ 128 $\times$ 128 grid points at the highest grid
level for the first three cases, and to 512 $\times$ 256 $\times$ 256
grid points for the last case. A SN explosion with energy $E_O = 10^{51}$ is
initially injected from the left-bottom corner of the box.

The numerical results are consistent with the previous analytical
results. Figure 4a depicts the initial interaction between a SNR
and a cloud with radius $r_c$ = 10 pc,  at a time t$ = 8 \times
10^4$ yr (which is comparable with the one estimated in the
analytical study). The initial conditions for this case represent
the cross in Figure 3 (bottom-left) which is outside of the shaded
area. The SNR shell in Figure 4a has a density $\sim$ 3.5 times
the ambient density and its velocity at the moment of the
interaction is of 240 km/s. The forward shock wave produced into the
cloud (Figure 4b) has a density $n_{c,sh} \sim$ 100 cm$^{-3}$, a
temperature $T_{c,sh} \sim 60$ K and a velocity $\hat{v}_{cs} \sim
5$ km s$^{-1}$. After a time of 3.7 $\times 10^6$ yr from the
start of the SNR-cloud interaction, the cloud is compressed in a cylindrical
core with a radius of $\sim$ 4 pc, an height of $\sim$ 3 pc, a density of
$\sim 220$ cm$^{-3}$ and temperature of $\sim 48$ K (Figure 4c). At this
point, the cloud core may collapse by its self-gravity or rebound
and start a re-expansion of the cloud material. In order to
collapse its mass should be larger than the Jeans mass. From Eq.
(22), this mass should be $M_j$ $\simeq$ 2140 M$_{\odot}$, while the
core mass inferred from the simulation is $M_c \sim 950$ M$_{\odot}$.
The cloud mass at the beginning of the simulation was $M_c = 1315$
M$_{\odot}$, and this
means that a mass $\sim 400$ M$_{\odot}$ has been dragged by the
SNR shell (see, e.g., the mass loss rate predicted by  Klein,
McKee \& Colella 1994), and that the shocked cloud is not
gravitationally unstable. After reaching a maximum density and
occupying a minimum volume, the gas begins to re-expand into the
ambient medium, and no dense structure develops (see Figure 4d).

We have also examined the cases of interactions between a SNR and
a cloud with a radius of 5 pc and 20 pc. In both cases, as
indicated by the crosses marked in Figure 3 (second and third
panels, respectively), the interactions are also unable to create
the conditions to produce a Jeans unstable cloud core. In the
first case ($r_c$=5 pc, Figure 5) the cloud is swept by the shock
front and no core survives after the interaction. In the second
case ($r_c$ = 20 pc, Figure 6), the shock wave has insufficient
energy to sweep all of the gas, and after a few Myrs it stalls
within the cloud with a velocity $\sim$ 0 (Figure 6, bottom-right
panel). We note that in this case the shocked cloud material does
not reach the conditions to become Jeans unstable, but the dense
cold shell that develops in the cloud may fragment and eventually
generate dense cores, as observed in most GMC.

Finally, we have also examined the interaction between a SNR and a
GMC having physical conditions which according to Figure 3 would
be able to generate a Jeans unstable core ($r_c$ = 10 pc, $n_c$ =
20 cm$^{-3}$, $R_{SNR} = 30$ pc) (see the cross within the shaded
area in Figure 3, bottom-left). This simulation was run with a
maximum resolution of 0.2 pc, that is twice of that adopted in the
other simulations. The evolution of the interacting system (Figure
7) indicates that after $\sim$ 4 Myr  a cold and dense cylindrical
core develops with radius of $\sim$ 3 pc, height of $\sim$ 3 pc,
density $n_{c,sh} \simeq$ 440 cm$^{-3}$, temperature $T_{c,sh}
\simeq 40$ K and a total mass of $\sim$ 1200 M$_{\odot}$. In this
case the Jeans mass (Eq. (22)) is $\sim$ 1000 M$_{\odot}$ and
therefore the conditions of the shocked gas are sufficient to form
a gravitationally unstable core. This  is in agreement with the
analytical results of Figure 3, but we note that in this
case, the distance between the SNR and the cloud is only 30 pc,
that is, much shorter than the proposed distance for the
SN-$\beta$ Pictoris system by Ortega et al. (2002, 2004).

\begin{figure}
\centering
\epsfxsize=9cm
\epsfbox{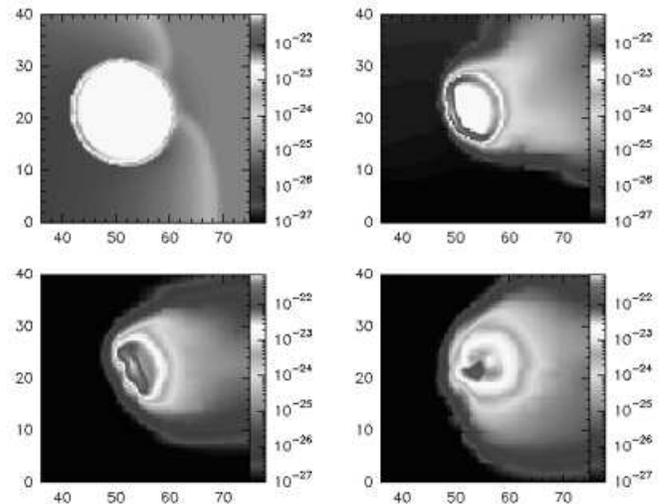}
\caption{Color-scale
maps of the midplane density distribution (in log scale) evolution
of the interaction between an expanding SNR and a cloud at a time
a) $t=2.2 \times 10^5$ yr (top-left); b) $t=2. \times 10^6$ yr
(top-right); c) $t=3.7 \times 10^6$ yr (bottom-left); and d)
$t=8.5 \times 10^6$ yr (bottom-right). The SNR is generated by a
SN explosion with an energy of $10^{51}$ erg. The ISM where the
SNR expands has a number density $n=0.05$ cm$^{-3}$, and a
temperature $10^4$ K. The cloud has an initial number density
$n_c=10$ cm$^{-3}$, temperature $T_c=$ 100 K, and radius $r_c=$ 10
pc. The initial distance between the external surface of the cloud
and the center of the SNR is 62 pc, as assumed by Ortega et al.
(2005).}
\end{figure}

\begin{figure}
\centering
\epsfxsize=9cm
\epsfbox{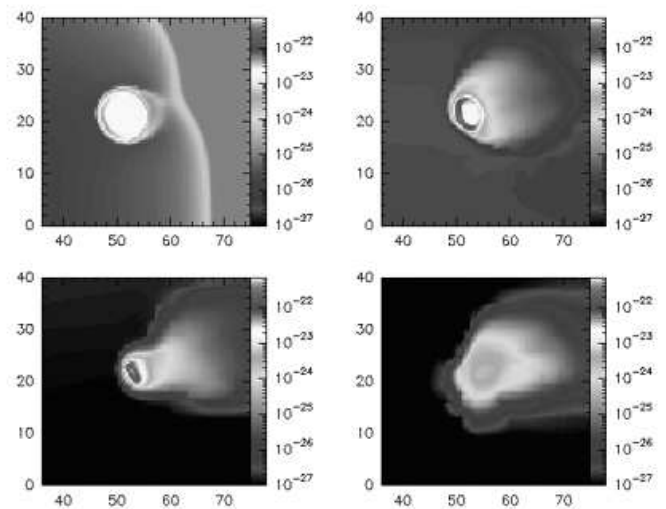}
\caption{The same as in
Figure 4, except that here the cloud has an initial radius $r_c$ =
5 pc. The times are:  $t=2.5 \times 10^5$ yr (top-left); $t=8.9
\times 10^5$ yr (top-right); $t=1.8 \times 10^6$ yr (bottom-left);
$t=8.5 \times 10^6$ yr (bottom-right).}
\end{figure}

\begin{figure}
\centering
\epsfxsize=9cm
\epsfbox{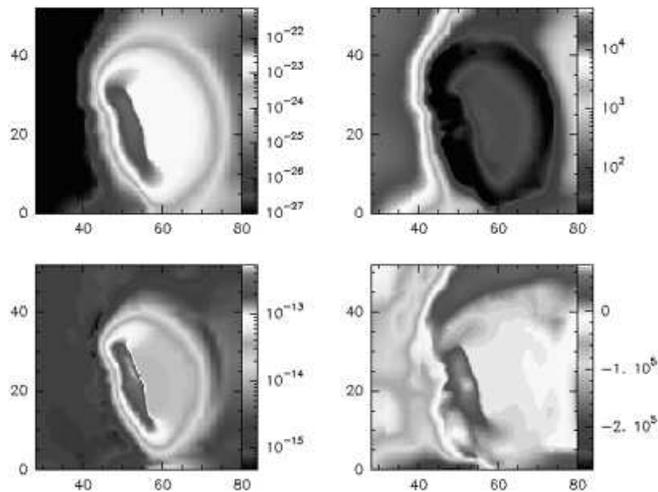}
\caption{Color log
scale maps of the midplane: a) density in cm s$^{-3}$(top-left);
b) temperature in K (top-right); c) pressure in dyne
(bottom-left); and velocity distribution in cm s$^{-1}$
(bottom-right) for a cloud with an initial radius $r_c$ = 20 pc at
a time $t=8 \times 10^6$ yr. The other initial conditions are the
same as in Figure 5.}
\end{figure}

\begin{figure}
\centering
\epsfxsize=9cm
\epsfbox{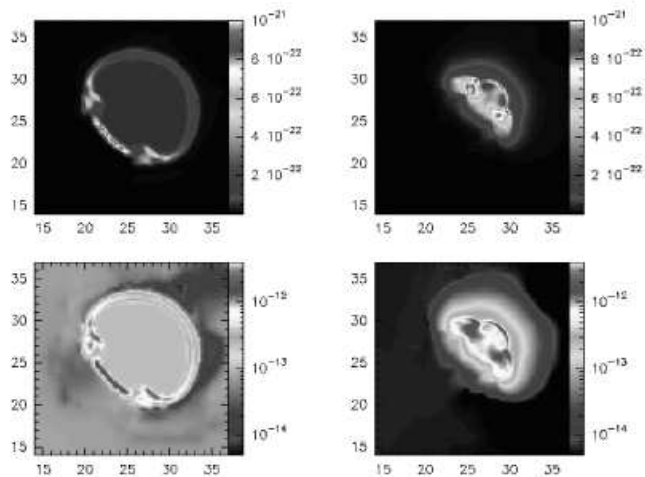}
\caption{Color log
scale maps of the midplane density (upper panels) and pressure
distributions (bottom panels) for a cloud with initial density
$n_c$ = 20 cm$^{-3}$ shocked by a SNR with $R_{SNR}=30$ pc. The
left panels are at a time $t=1.6 \times 10^6$ yr and the right
panels are at $t=4.1 \times 10^6$ yr.}
\end{figure}

\section{Conclusions}

We have here presented a preliminary study of the role that
interactions between SNRs and dense molecular clouds play in the process of
star formation.

Considering the physical conditions that are relevant  in these
interactions for triggering star formation (like the choice of the
cloud and SNR initial conditions, the derivation of an appropriate
Jeans mass, the determination of the conditions for complete
destruction of the cloud after the impact, and the determination
of the extent of penetration of the shock front into the cloud) we
have built in Section 3, a diagram of the SNR radius versus the
cloud density for a fixed cloud radius in which the constraints
above delineate a shaded zone where star formation induced by SN
shock front-cloud interactions is possible.

We have also performed fully 3-D radiatively cooling numerical
simulations of the impact between a SNR and a cloud for different
initial conditions (in Section 4) and, although self-gravity has
not been included in the present study, we have been able to track
the first steps of these interactions and detect the conditions
that lead  either to cloud collapse and star formation or to
complete cloud destruction. We have found that the  numerical
results are consistent with those established by the SNR-cloud
density diagram in spite of the fact that the later
have been built from approximate conditions derived
from a simplified analytic theory.

Finally, we have applied the results above to the nearby young
stellar association $\beta-$ Pictoris which is composed of low mass
Post-T Tauri stars with an age of 11 Myr. Ortega et al. (2004)
have recently suggested that its formation could have been
triggered by the shock wave produced by a SN explosion  that may
have occurred either in the Lower Centaurus Crux (LCC) or in Upper
Centaurus Lupus (UCL) older subgroups of the OB Cen-Sco
association.  
Taking from their study the initial conditions that
would be appropriate for both the ISM and the cloud and assuming
an interacting SN shock front still in the adiabatic phase, we
have found that the suggested  origin for the young association by
Ortega et al. is implausible, for the proposed distance of
the SNR, of  $\sim 60$ pc, unless the parent molecular cloud had a
radius of the order of 10 pc and a density of the order of 20
$cm^{-3}$, as indicated by the shaded zone of the third panel of
Figure 3. A larger cloud radius ($\sim 20$ pc) would require a
much smaller cloud density (see bottom panel of Figure 3) which
would not be much frequent in GMCs with a temperature greater than 50 K. 
These analytical results have also been confirmed by the numerical
simulations (see Figures 5 to 7). The results indicate that,
unless the SN had an extremely (unusual) high energy, or had
exploded at a much smaller distance than the proposed one (of
$\sim 60$ pc), then the suggested scenario for triggering the
formation of the $\beta-$ Pictoris association would be possible
only under the restrict conditions above. In fact, our numerical
results also revealed that using a similar parent cloud density to
that proposed above, but assuming an interaction with a SNR at
half the distance (30 pc) and a cloud with larger radius (20 pc)
the interaction may lead to the development of structuring and
formation of dense cold clumps behind the shock in the cloud,
which may eventually aggregate and generate dense Jeans unstable
cores, as observed in most GMCs.

We should emphasize that the present study is still preliminary
and the results have been mainly focused on the particular
conditions of the $\beta$ Pictoris association and its neighbourhood.
Besides, it has been performed without taking into account the
presence of magnetic fields. These may also play
an important role on star formation as they tend to attain
values as large as $\sim  100 \mu$G to few mG within the densest
cores of the molecular clouds and can dominate over the thermal and
turbulent pressures, so that  at least the densest clouds may be
magnetically supported. A more general investigation of star
formation processes induced by SNR shock fronts including not only
radiative cooling but also magnetic fields and self-gravity is in
preparation and will be presented elsewhere (Melioli et al. 2006)

\section*{Acknowledgments}
C.M. and E.M.G.D.P acknowledge financial support from the
Brazilian Agencies FAPESP and CNPq. The authors are also indebted to an 
anonymous referee for his/her careful revision and 
comments on this work

{}

\bsp

\label{lastpage}


\begin{thebibliography}{99}

\bibitem[]{}
Blitz, L., 1993,  Protostars and planets III, p. 125
\bibitem[]{}
Blitz, L., Williams, J.P., 1999, The Origin of Stars and Planetary Systems.
Edited by C.J. Lada \& N.D. Kylafis. Kluwer Acad. Publ., p.3
\bibitem[]{}
Bonnell, I.A., Dobbs, C.L., Robitaille, T.P. \& Pringle, J.E., 2006, MNRAS,
365, 37
\bibitem[]{}
Dalgarno, A.\& McCray, R.A., 1972, ARA\&A, 10, 375
\bibitem[]{}
de la Reza, R., Jilinski, G. \& Ortega, V.G. 2006 to appear in the
AJ (May)
\bibitem[]{}
Elmegreen, B.G., \& Scalo, J., 2004, ARA\&A, 42, 21
\bibitem[]{}
Hoogerwerf, R., de Bruijne, J.H.J. \& de Zeeuw, P.T. 2001 A\&A 365,
49
\bibitem[]{}
Jeans, J.H., 1902, Phil. Trans. A., 199, 1
\bibitem[]{}
Klein R., McKee C.F. \& Colella P., 1994, ApJ, 420, 213
\bibitem[]{}
Kornreich, P. \& Scalo, J., 2000, ApJ, 531, 366
\bibitem[]{}
Larson, R.B., 1981, MNRAS, 194, 809
\bibitem[]{}
Mac Low, M-M., \& Klessen, R.S., 2004, RvMP, 76, 125
\bibitem[]{}
Masciadri E., de Gouveia Dal Pino E.M., Raga A.C. \& Noriega-Crespo A., 2002,
ApJ, 580, 950
\bibitem[]{}
McCray R., 1985, in Spectroscopy of Astrophysical Plasmas, edited by A.
Delgarno \& D. Layzer, p270
\bibitem[]{}
Melioli, C., de Gouveia Dal Pino, E.M., \& Raga, A., 2005, A\&A,
443, 495
\bibitem[]{}
Melioli, C., de Gouveia Dal Pino, E.M., Le\~ao M. R. M., \& Raga,
A., 2006, in prep.
\bibitem[]{}
Nakamura, F., McKee, C.F., Klein, R.I., Fisher, R.T., 2005, astro-ph/0511016
\bibitem[]{}
Ortega, V.G., de la Reza, R., Jilinski, E. \& Bazzanella, B., 2004, ApJ,
609, 243
\bibitem[]{}
Ortega, V.G., de la Reza, R., Jilinski, E. \& Bazzanella, B., 2002, ApJ,
575, 75
\bibitem[]{}
Poludnenko A.Y., Frank A. \& Blackman E.G., 2002, ApJ, 576, 832
\bibitem[]{}
Raga A.C., de Gouveia Dal Pino E.M., Noriega-Crespo A., Minnini P.D. \&
Vel\'azquez P.F., 2002, A\&A, 392, 267
\bibitem[]{}
Raga A.C., Navarro-Gonz\'alez, R., \& Villagr\'an-Muniz, M. 2000, Rev. Mexicana
Astron. Astrofis., 36, 67
\bibitem[]{}
Roberts, W.W., 1969, ApJ, 158, 123
\bibitem[]{}
Torres, C.A.O., Quast, G. R., da Silva, L., de la Reza, R., Melo,
C.H.F. \& Sterzik, M. 2006 to be submitted to the A\&A
\bibitem[]{}
Wada, K. \& Norman, C.A., 2001, ApJ, 547, 172
\bibitem[]{}
Williams, J.P., Blitz, L., McKee, C.F., 2000, Protostars and Planets IV,
eds Mannings, V., Boss, A.P., Russell, S.S., p. 97
\bibitem[]{}
Zuckerman, B., Song, I., Bessell, M.S. \& Webb, R.A., 2001, ApJ, 562, 87

\end{thebibliography}
\end{document}